\documentclass[11pt,a4paper]{article}

\pdfoutput=1

\usepackage{jcappub}

\usepackage{epsfig}
\usepackage{dcolumn}
\usepackage{bm}
\usepackage{amssymb}
\usepackage{amsmath}
\usepackage{amsfonts}
\usepackage{graphicx}
\usepackage{psfrag}

\def\R{{\cal R} }
\def\k{{\mathbf{k}}}
\def\p{{\mathbf{p}}}
\numberwithin{equation}{section}

\subheader{\begin{flushright}
UTTG-23-11\\
TCC-027-11
\end{flushright}}

\title{Inflation with general initial conditions for scalar perturbations}

\author{Sandipan Kundu}

\affiliation{Texas Cosmology Center, University of Texas, Austin, TX 78712}
\affiliation{Theory Group, Department of Physics, University of Texas, Austin, TX 78712}

\emailAdd{sandyk@physics.utexas.edu}

\abstract{We explore the possibility of a single field quasi-de Sitter inflationary model with general initial state for primordial fluctuations. In this paper, first we compute the power spectrum and the bispectrum of scalar perturbations with coherent state as the initial state. We find that a large class of coherent states are indistinguishable from the Bunch-Davies vacuum state and hence consistent with the current observations. In case of a more general initial state built over Bunch-Davies vacuum state, we show that the constraints on the initial state from observed power spectrum and local bispectrum are relatively weak and for quasi-de Sitter inflation a large number of initial states are consistent with the current observations.  However, renormalizability of the energy-momentum tensor of the fluctuations constraints the initial state further.}

\arxivnumber{1110.4688}

\begin{document}
\maketitle
\flushbottom

\section{Introduction}
Inflation has been very successful in explaining several puzzles of the standard big bang scenario. But the most important success of the inflationary theory, is its prediction of almost scale invariant power spectrum of primordial fluctuations\cite{Starobinsky:1982ee, Hawking:1982cz, Guth:1982ec, Bardeen:1983qw, Mukhanov:1985rz}. In the inflationary scenario the temperature fluctuations of cosmic microwave background (CMB) and the large scale structure (LSS) are directly related to the curvature perturbations produced during inflation. Current observations of CMB strongly support the presence of an almost scale-invariant power spectrum.

From a theoretical point of view, inflation is important because it gives an opportunity to test predictions of quantum field theory in a curved space time. In this framework, the choice of vacuum for quantum fluctuations is ambiguous.  Hamiltonian for the primordial fluctuations is time-dependent and hence the meaning of vacuum is not very clear. Power spectrum of primordial fluctuations is generally computed assuming that the fluctuations are initially in the Bunch-Davies vacuum state\cite{Bunch:1978yq,Birrell:1982ix}. Bunch-Davies vacuum state is the minimum energy eigenstate of the Hamiltonian in the infinite past and it is a reasonable choice as an initial state but not unique. It is somewhat of a philosophical question whether initial conditions are integral part of a theory or should be analyzed separately. There has been a great deal of work focused on modifications of Bunch-Davies vacuum state\cite{Kaloper:2002cs, Greene:2005aj, Easther:2005yr, Brunetti:2005pr, Goldstein:2003ut, Collins:2003mj, Collins:2003zv, Danielsson:2002kx, Danielsson:2004xw, Giovannini:2010xg, Dey:2011mj, Carney:2011hz}. In this paper, for simplicity we will assume that the Bunch-Davies state represents an empty vacuum state in the beginning of inflation. Even with that, there is no good reason why the initial state has to be a vacuum state. We do not know anything about the physics before inflation and a priori any excited state is as good an initial state as the vacuum state. In fact excited initial states can open interesting possibilities. We also believe understanding the initial state of the primordial fluctuations will be the first step towards the physics before inflation.

In spite of the great success of the inflationary theory, it is always important to verify the validity of different assumptions. The purpose of this paper is to discuss the effect of relaxing the assumption of Bunch-Davies vacuum by choosing a general initial state built over the Bunch-Davies vacuum state. The effect of having thermal initial state on the power spectrum has already been considered in \cite{Bhattacharya:2005wn, Ferreira:2007cb}.\footnote{See \cite{Gasperini:1993yf} for a discussion on graviton spectrum from a thermal initial state.} In \cite{Ashoorioon:2010xg} authors have identified a family of excited states that are indistinguishable from the Bunch-Davies vacuum  at the level of two-point function or three-point function. Similar work has also been carried out by Agullo and Parker in \cite{Agullo:2010ws} but our approach in this paper is different. Today our universe is homogeneous and isotropic on the large scale and the power spectrum of primordial fluctuations is nearly scale-invariant from the size of the observable universe down to the scales of around a Mpc; as the authors show in \cite{Peiris:2009wp}. Also the measurements of the bispectrum (three-point function) from the CMB\cite{Komatsu:2010fb} and LSS\cite{Slosar:2008hx} indicate that the primordial fluctuations are nearly gaussian. The current observational constraint on the non-gaussianity parameter $f_{NL}$ is very weak and $f_{NL}^{loc}$ remains the best constrained non-gaussianity parameter: $-5 <f_{NL}^{loc}<59$ (WMAP7+SDSS). Quasi-de Sitter inflation (or slow roll inflation) with Bunch-Davies vacuum state as the initial state for the fluctuations produces an almost scale-invariant power spectrum ($n_s\approx 1$) and a negligible non-gaussianity ($f_{NL}^{loc}\approx 0.02$)\cite{Maldacena:2002vr}. As a first step, we will compute the power spectrum and the bispectrum with coherent state as the initial state and show that a large class of coherent states can produce exactly the same power spectrum and bispectrum as the vacuum state. 

We will then compute the power spectrum for quasi de Sitter inflation with a general initial state and find the constraint on the initial state by demanding scale invariance. Constraints on the initial state from current measurements of power spectrum are relatively weak. For quasi-de Sitter inflation, Bunch-Davies vacuum and coherent states are not the only states, in fact a large number of states are consistent with the observations. However, we will demonstrate that renormalizability of the energy-momentum tensor of the primordial fluctuations imposes some restrictions on the initial state. We will also argue that $f_{NL}^{loc}\approx\frac{5}{12}(1-n_s)$ holds even for a general initial state. Therefore,  if a state  produces an almost scale-invariant power spectrum then it can not produce large $f_{NL}^{loc}$.

The rest of the paper is organized as follows. In section \ref{review}, we review the quantization of the fluctuations in inflationary universe. In section \ref{coherentstate}, we compute the power-spectrum and the  bispectrum with coherent state. Section \ref{genstate} is devoted to the introduction of excited initial states and constraints on the initial state from observations. In section \ref{renormalization}, we discuss renormalizability of the energy-momentum tensor of the fluctuations and constraints that imposes on the initial state. We end with a discussion on bispectrum in squeezed limit with a general state in section \ref{bispectrum} and concluding remarks in section \ref{conclusions}.

\section{Review of quantization of the fluctuations in inflationary universe}\label{review}

\subsection{Scalar field in FRW universe}
We start with the Lagrangian of gravity and minimally coupled real scalar field (for a nice review see \cite{Baumann:2009ds})

\begin{equation}\label{EHaction}
 S=\frac{1}{2}\int \sqrt{-g}d^4x\left[\frac{1}{8\pi G}R - g^{\mu\nu}\partial_\mu \phi \partial_\nu \phi -2V(\phi)\right].
\end{equation}
A homogeneous background solution has the form
\begin{equation}
 ds^2= -dt^2+ a^2(t)d\mathbf{x}^2
\end{equation}
with a background scalar field $\phi(\mathbf{x}, t)= \bar{\phi}(t)$. This background obeys the equations
\begin{align}
& 3H^2= 8 \pi G\left[\frac{1}{2}\dot{\bar{\phi}}^2 +V(\bar{\phi})\right],\\
&\ddot{\bar{\phi}}+3H\dot{\bar{\phi}}+V'(\bar{\phi})=0,
\end{align}
where $H$ is the Hubble parameter $H=\dot{a}/a$. Next we consider perturbations around the homogeneous background solutions. 
\begin{equation}
  \phi(\mathbf{x}, t)= \bar{\phi}(t)+ \delta \phi(\mathbf{x}, t)
\end{equation}
and the perturbed metric with only scalar perturbations is 
\begin{equation}
 ds^2=-(1+2\Phi)dt^2+2a(t)(\partial_{i}B) dx^idt+a^2(t)[(1-2\Psi)\delta_{ij}+2\partial_{ij}E]dx^idx^j.
\end{equation}
We can avoid fictitious gauge modes by introducing gauge-invariant variables\cite{Bardeen:1980kt, Lyth:1984gv}. One such variable is the comoving curvature perturbation 
\begin{equation}\label{curvpert}
\R= \Psi+\frac{H}{\dot{\bar{\phi}}}\delta\phi.
\end{equation}
Expanding the action (\ref{EHaction}), we get the second order action for the scalar fluctuations in terms of the gauge-invariant variable $\R$
\begin{equation}\label{secondaction}
S_{2}=\frac{1}{2} \int d^4x a^3 \frac{\dot{\bar{\phi}}^2}{H^2}\left[\dot{\R}^2-a^{-2}(\partial_i \R)^2 \right].
\end{equation}
The next step is to quantize this system. Before doing that to simplify the action, we introduce Mukhanov variable $v=z \R$ and conformal time $\tau$, where $z=a \dot{\bar{\phi}}/H$. In terms of the equation of state parameter, $z^2 = \frac{3}{8\pi G} a^2 (1 + w)$.\footnote{Equation of state parameter is defined by $p=w \rho$, where $p$ is the pressure and $\rho$ is the energy density.} That leads to the action 
\begin{equation}\label{mukhanovaction}
S_{2}=\frac{1}{2} \int d\tau d^3x \left[(v')^2- (\partial_i v)^2 + v^2 b(\tau)\right],
\end{equation}
where $(...)'=\partial_\tau(...)$ and $b(\tau)= \left(z''/z\right)$. We can define the Fourier transform of the $v$ field in the standard way,
\begin{equation}
v(\mathbf{x}, \tau)=\int \frac{d^3\mathbf{k}}{(2\pi)^{3}}  v_{\mathbf{k}}(\tau)e^{i \k.\mathbf{x}} 
\end{equation}
with $v^*_{\k}(\tau)=v_{-\k}(\tau)$ (because field $v$ is real). And in terms of the Fourier transform the action (\ref{secondaction}) now reads
\begin{equation}
S_2=\frac{1}{2}\frac{1}{(2 \pi)^3}\int d\tau d^3\k \left[v'_{\k}(\tau){v_{\k}^{*}}'(\tau)-k^2 v_{\k}(\tau)v^{*}_{\k}(\tau)+b(\tau)v_{\k}(\tau)v^{*}_{\k}(\tau)\right]. 
\end{equation}
From this action we get the following equation\cite{Mukhanov:1985rz, Mukhanov:1990me} for $v_{\k}$
\begin{equation}\label{mukhanovequation}
v_{\k}''(\tau)+ \omega_k^2(\tau) v_{\k}(\tau)=0 
\end{equation}
with $ \omega_k^2(\tau)= k^2-b(\tau)$. If $u_k(\tau)$ and $u_k^*(\tau)$ are linearly independent and form a basis in the space of complex solutions of equation (\ref{mukhanovequation}) then the Wronskian $W[u_k,u_k^*]=2 i Im[u'_k(\tau) u^*_k(\tau)]\neq 0$. Complex solution $u_k(\tau)$ is called the mode function. Wronskian $W$ is time-independent so we can always normalize the mode function $u_k(\tau)$ by the condition 
\begin{equation}\label{normalization}
Im[u'_k(\tau) u^*_k(\tau)]=1. 
\end{equation}
The general solution of equation (\ref{mukhanovequation}) can be written as
\begin{equation}\label{generalsolution}
v_{\k}(\tau)=\frac{1}{\sqrt{2}}\left[a^-_{\k}u^*_k(\tau)+ a^+_{-\k}u_k(\tau)\right],
\end{equation}
where $a^-_{\k}$ and $a^+_{-\k}$ are independent of $\tau$ and $ a^+_{\k}= (a^-_{\k})^*$ (because $v(\mathbf{x}, \tau)$ is real).

\subsection{Quantization of fluctuations}
Field $v(\mathbf{x}, \tau)$ can be quantized in the standard way, just like harmonic oscillator (we will work in the Heisenberg picture). We introduce the commutation relation
\begin{equation}\label{commutation1}
\left[\hat{v}(\mathbf{x}, \tau), \hat{\pi}(\mathbf{y}, \tau)\right]=i \delta^3(\mathbf{x}-\mathbf{y}), 
\end{equation}
where $\hat{\pi}= \hat{v}'$ is the canonical momentum. Now the equation (\ref{generalsolution}) becomes
\begin{equation}\label{quantumv}
 \hat{v}_{\k}(\tau)=\frac{1}{\sqrt{2}}\left[\hat{a}_{\k}u^*_k(\tau)+ \hat{a}^\dagger_{-\k}u_k(\tau)\right].
\end{equation}
Now comoving curvature perturbation is an operator $\hat{\R}=\hat{v}/z$. The commutation relation (\ref{commutation1}) gives commutation relations between $\hat{a}^\dagger_{\k}$ and $\hat{a}_{\k}$ 
\begin{equation}\label{commutation}
\left[\hat{a}_{\k_1},\hat{a}^\dagger_{\k_2}\right]=(2 \pi)^3 \delta^3(\k_1-\k_2),  \qquad   \left[\hat{a}^\dagger_{\k_1},\hat{a}^\dagger_{\k_2}\right]=\left[\hat{a}_{\k_1},\hat{a}_{\k_2}\right]=0 .
\end{equation}
The Hamiltonian of the system is
\begin{equation}\label{hamiltonian}
\hat{H}(\tau)=\frac{1}{4}\frac{1}{(2 \pi)^3}\int d^3\k \left[\hat{a}_{\k}\hat{a}_{-\k} F_k^*(\tau) + \hat{a}^\dagger_{\k}\hat{a}^\dagger_{-\k} F_k(\tau) + \left(\hat{a}_{\k}\hat{a}^\dagger_{\k}+\hat{a}^\dagger_{\k}\hat{a}_{\k}\right) E_k(\tau)\right],
\end{equation}
where,
\begin{equation}
F_k(\tau)= (u_k')^2+ \omega_k^2 u_k^2,  \qquad   E_k(\tau)= |u_k'|^2+ \omega_k^2 |u_k^2|.
\end{equation}

\subsection{Bunch-Davies vacuum}
The next step is to define a vacuum state and find out the mode-functions that describe the state. The Hamiltonian explicitly depends on the conformal time $\tau$. So it is not possible to define a vacuum in a time-independent way. First we define the vacuum by the standard condition: for all $\k$
\begin{equation}\label{vaccond1}
\hat{a}_{\k}|0\rangle=0 . 
\end{equation}
But this is not enough to specify the mode-functions. It is not possible to find a time-independent eigenstate of the Hamiltonian. So we take a particular moment $\tau=\tau_0$, and define the vacuum as the lowest-energy eigenstate of the instantaneous Hamiltonian of the fluctuations at $\tau=\tau_0$ (we can always do that as long as $\omega_k^2(\tau_0)\geq0$). That gives us following initial conditions for the mode functions 
\begin{equation}\label{initialcondition}
u_k'(\tau_0)=\pm i \sqrt{\omega_k(\tau_0)}e^{i \lambda(k)},  \qquad  u_k(\tau_0)=\pm  \frac{1}{\sqrt{\omega_k(\tau_0)}}e^{i \lambda(k)},
\end{equation}
where $\lambda(k)$ is some arbitrary time independent function of $k$. In the limit when $\tau_0$ represents infinite past (i.e. $\tau_0\rightarrow -\infty$), the vacuum is called Bunch-Davies vacuum state. In this limit, $\omega_k^2=k^2\geq0$ and we can define vacuum by equation(\ref{vaccond1}) for all modes.

\subsection{Power-spectrum for inflation}
During inflation,\footnote{Here we will assume that $V(\bar{\phi})$ is approximately constant and slow roll parameters $|\epsilon|,|\eta |\ll 1$, where
\begin{align}\label{slowroll}
\epsilon=&-\frac{\dot{H}}{H^2}=\frac{3}{2}(w+1)\approx\frac{1}{16 \pi G}\left(\frac{V'}{V}\right)^2 ,\\
\eta=&\frac{1}{8 \pi G}\frac{V''}{V}.
\end{align}
It is easy to check that $H$ and $\dot{\bar{\phi}}$ are also approximately constants.} we have $w\approx-1$ (see \cite{Ackerman:2010he} for constraints on $w$) and $b(\tau)= (2/\tau^2)$. Solving equation (\ref{mukhanovequation}) with normalization condition (\ref{normalization}) and initial conditions (\ref{initialcondition}) (with the $+$ sign and $\tau_0\rightarrow -\infty$ limit), we get 
\begin{equation}\label{modesBD}
u_k(\tau)= \frac{e^{ik\tau}}{\sqrt{k}}\left(1+\frac{i}{k\tau}\right). 
\end{equation}
With this mode functions we can now compute the power-spectrum. For that we need to calculate following quantity
\begin{equation}\label{vk2}
\langle\hat{v}_{\k}(\tau)\hat{v}_{\k'}(\tau)\rangle\equiv\langle 0|\hat{v}_{\k}(\tau)\hat{v}_{\k'}(\tau)|0\rangle=\frac{1}{2} (2 \pi)^3 \delta^3(\k+\k')|u_k(\tau)|^2, 
\end{equation}
where we got the last equation using (\ref{quantumv}). Before we compute $\langle\hat{\R}_{\k}(\tau)\hat{\R}_{\k'}(\tau)\rangle$, we should introduce some standard quantities
\begin{equation}\label{ns}
\langle\hat{\R}_{\k}(\tau)\hat{\R}_{\k'}(\tau)\rangle=(2\pi)^3\delta^3(\k+\k')P_{\R},  \qquad  \Delta^2_{\R}=\frac{k^3}{2\pi^2}P_{\R}, \qquad n_s-1=\frac{d\ln \Delta^2_{\R}}{d \ln k} ,
\end{equation}
where $n_s$ is called the scalar spectral index or tilt. Using equations (\ref{modesBD}-\ref{ns}), in the superhorizon limit ($|k\tau|\ll1$),we have
\begin{equation}
n_s=1,  \qquad   \Delta^2_{\R}=\frac{ H^4}{4\pi^2~ \dot{\bar{\phi}}^2 }.
\end{equation}
Where $H$ is the Hubble parameter during inflation. Observation of CMB and LSS tells us conclusively that the power spectrum of the fluctuations produced during inflation is almost scale-invariant (i.e. $n_s\approx 1$). Here we have ignored the slow roll parameters because deviation from perfect scale invariance is small. In the first order in slow roll parameters, spectral index is given by 
\begin{equation}\label{slowrollns}
n_s= 1-6\epsilon +2\eta.
\end{equation}

\section{Coherent states}\label{coherentstate}
In this section, we will compute the power spectrum and the bispectrum with a non-vacuum initial state namely, coherent state. Coherent state is a special state because it closely resembles classical harmonic oscillation. But the obvious question is: why coherent state? Because we do not know anything about the physics before inflation, a priori any excited state is as good an initial state as the vacuum state. In particular, coherent state is an interesting example of a non-vacuum state and we will show that a large number of coherent states are indistinguishable from the Bunch-Davies vacuum state. And it is a good first step towards a better understanding of the initial condition of primordial fluctuations.

We can use $\hat{a}^\dagger_\k$ operators to build excited states over the Bunch-Davies vacuum state $|0\rangle$ 
\begin{equation}\label{excitedstate}
|\psi\rangle= \frac{1}{\sqrt{n_1!n_2!...}} \left[\left(\hat{a}^\dagger_{\k_1}\right)^{n_1}\left(\hat{a}^\dagger_{\k_2}\right)^{n_2}...\right]|0\rangle. 
\end{equation}
The coherent state is defined in the usual way,
\begin{equation}\label{coherent}
\hat{a}_{\k}|C\rangle=C(\k)|C\rangle 
\end{equation}
with $\langle C|C\rangle=1$. Now the vacuum state is the special case: $C(\k)=0$. 


\subsection{Power-spectrum with coherent states}
Before marching onwards with our calculations, we want to impose the following constraint on the initial state. We want $\langle\hat{\R}_{\k}(\tau)\rangle=0$, in superhorizon limit. That leads to the condition
\begin{equation}\label{concoherent}
C^{*}(-\k)=C(\k). 
\end{equation}
It is important to note that renormalizability of energy-momentum tensor of the fluctuations (see section \ref{renormalization}) constraints $C(\k)$ further: $C(\k)$ goes to zero faster than $\frac{1}{k^{5/2}}$ for large $\k$.

Now we can compute the power spectrum 
\begin{equation}
\langle\hat{\R}_{\k}(\tau)\hat{\R}_{\k'}(\tau)\rangle=\langle C|\hat{\R}_{\k}(\tau)\hat{\R}_{\k'}(\tau)|C\rangle ,
\end{equation}
using 
\begin{equation}
\hat{\R}_{\k}(\tau)=\frac{1}{\sqrt{2}}\left[\hat{a}_{\k}\R^*_k(\tau)+ \hat{a}^\dagger_{-\k}\R_k(\tau)\right]
\end{equation}
where, $\R_k(\tau)=\left(\frac{H}{a \dot{\bar{\phi}}} \right) \frac{e^{ik\tau}}{\sqrt{k}}\left(1+\frac{i}{k\tau}\right)$. With the help of equations(\ref{coherent},\ref{concoherent}) and commutation relations (\ref{commutation}) in the superhorizon limit ($|k \tau|,|k' \tau|\ll 1$), we get the following result
\begin{equation}\label{powerspeccoherent}
\langle\hat{\R}_{\k}(\tau)\hat{\R}_{\k'}(\tau)\rangle=\frac{1}{2} (2 \pi)^3 \frac{ H^4}{ \dot{\bar{\phi}}^2 k^3}\delta^3(\k+\k').
\end{equation}
Therefore, spectral index $n_s=1$.\footnote{If we don't neglect the slow roll parameters then the spectral index, just like equation (\ref{slowrollns}), is given by $n_s= 1-6\epsilon +2\eta$.}  Note that equation(\ref{powerspeccoherent}) is exactly the same as the power spectrum with Bunch-Davies vacuum state. It is not very difficult to understand why we have the same result. One can think of coherent state as the zero-point quantum fluctuations around some classical state. So, the field $\hat{\R}^{coh}_{\k}(\tau)$ in the coherent state can be written as $\hat{\R}^{coh}_{\k}(\tau)=\R^{cl}(\tau) +\hat{\R}^{vac}_{\k}(\tau)$, where classical part $\R^{cl}(\tau)$ is obviously the expectation value $\langle\hat{\R}^{coh}_{\k}(\tau)\rangle$. And $\hat{\R}^{vac}_{\k}(\tau)$ is the original quantum field but now in the vacuum state. Because of (\ref{concoherent}), it is very easy to check that the classical part $\R^{cl}(\tau)$ decays and  in the superhorizon limit $\R^{cl}(\tau)$ vanishes. So, finally we end up with $\langle\hat{\R}^{coh}_{\k_1}(\tau)\hat{\R}^{coh}_{\k_2}(\tau)\rangle=\langle\hat{\R}^{vac}_{\k_1}(\tau)\hat{\R}^{vac}_{\k_2}(\tau)\rangle$.


\subsection{Primordial non-gaussianities from coherent state}
Non-gaussianity (for a good review see \cite{Bartolo:2004if}) is emerging as a very important tool for discriminating between different models of inflation. Even for slow roll inflation three-point function may contain more information about the initial state. As we have seen already that coherent state produces scale-invariant power spectrum just like the vacuum state. In this section we will compute the three-point function with coherent state (with slow-roll approximations) to see if there is any observable signature. To do that we will follow an approach similar to \cite{Agullo:2010ws} by Agullo and Parker. 
At leading order in the slow-roll parameters, the third order action for the scalar fluctuations is given by \cite{Maldacena:2002vr}
\begin{equation}\label{thirdorder}
S_3= 8 \pi G \int d^3x d\tau ~ a^3(\tau)\left( \frac{\dot{\bar{\phi}}}{H}\right)^4 H \R_c'^2 \partial ^{-2}\R_c',
\end{equation}
where, $\R_c$ is the redefined field
\begin{equation}
\R=\R_c +\frac{1}{4} \left(3\epsilon -2\eta\right)\R_c^2+ \frac{1}{2} \epsilon ~ \partial ^{-2}\left(\R_c \partial^2 \R_c\right).
\end{equation}
In momentum space the last equation becomes
\begin{equation}\label{rc}
\R_{\k}=\R_{c,\k}+\frac{1}{4}\left(3\epsilon -2\eta\right)\int \frac{d^3\mathbf{\p}}{(2\pi)^{3}}\R_{c,\p}\R_{c,\k-\p}+\frac{1}{2} \epsilon ~\int \frac{d^3\mathbf{\p}}{(2\pi)^{3}}\frac{(\k-\p)^2}{k^2}\R_{c,\p}\R_{c,\k-\p}.
\end{equation}
The interaction Hamiltonian can be found from $S_3=- \int d\tau H_{int}$. In momentum space $H_{int}$ is given by
\begin{equation}\label{Hint}
H_{int}(\tau)=\frac{8 \pi G}{(2\pi)^6}  a^3(\tau)\left( \frac{\dot{\bar{\phi}}}{H}\right)^4 H \int d^3\p_1 d^3\p_2 d^3\p_3 \left(\frac{1}{p_3^2}\right) \R'_{\p_1}(\tau)\R'_{\p_2}(\tau)\R'_{\p_3}(\tau) \delta^3(\p_1+\p_2+\p_3).
\end{equation}
We will use time dependent perturbation theory to compute the momentum space three point function. But before that it is important to note that in the presence of interaction (\ref{Hint}), $\langle C|\hat{\R}_{\k}(\tau\rightarrow 0)|C\rangle\neq 0$. So the quantity we are interested in is not $\langle C|\hat{\R}_{\k_1}(\tau)\hat{\R}_{\k_2}(\tau)\hat{\R}_{\k_3}(\tau)|C\rangle$ but $\langle C|\hat{\R}^{phy}_{\k_1}(\tau)\hat{\R}^{phy}_{\k_2}(\tau)\hat{\R}^{phy}_{\k_3}(\tau)|C\rangle$, where  $\hat{\R}^{phy}_{\k}(\tau')\equiv \hat{\R}_{\k}(\tau')-\langle C|\hat{\R}_{\k}(\tau\rightarrow 0)|C\rangle$. Therefore, at leading order in slow-roll parameters,\footnote{ $\langle C|\hat{\R}_{\k}(\tau)|C\rangle= 0$ for the Bunch-Davies vacuum and hence $\hat{\R}^{phy}_{\k}(\tau')=\hat{\R}_{\k}(\tau')$.}
\begin{align}\label{effthreepointfun}
\langle C|\hat{\R}^{phy}_{\k_1}(\tau)\hat{\R}^{phy}_{\k_2}(\tau)\hat{\R}^{phy}_{\k_3}(\tau)|C\rangle=&\langle C|\hat{\R}_{\k_1}(\tau)\hat{\R}_{\k_2}(\tau)\hat{\R}_{\k_3}(\tau)|C\rangle \nonumber \\
&-\left(\langle C|\hat{\R}_{\k_1}(\tau)|C\rangle \langle C|\hat{\R}_{\k_2}(\tau)\hat{\R}_{\k_3}(\tau)|C\rangle+ \text{cyclic perm}\right).
\end{align}
Using time-dependent perturbation theory we can calculate the three-point function (see Appendix \ref{appendix} for explicit calculations) and in the limit $\tau\rightarrow 0$, the final answer is
\begin{align}\label{resultnongauss}
\langle C|\hat{\R}^{phy}_{\k_1}(\tau)\hat{\R}^{phy}_{\k_2}(\tau)\hat{\R}^{phy}_{\k_3}(\tau)&|C\rangle=(2 \pi)^3 \delta^3(\k_1+\k_2+\k_3)P_R(k_2)P_R(k_1)\nonumber \\
&\times \left[\frac{1}{2}\left(3\epsilon -2\eta+\epsilon\frac{k_1^2+k_2^2}{k_3^2}\right)+\frac{4\epsilon }{ (k_1+k_2+k_3)}\frac{k_1^2k_2^2}{k_3^3}\right]+ \text{cyclic perm}.
\end{align}
Surprisingly, the last result does not contain the function $C(\k)$ and the result is exactly the same as the result for the vacuum state
\begin{equation}\label{nongausscoh}
\langle C|\hat{\R}^{phy}_{\k_1}(\tau)\hat{\R}^{phy}_{\k_2}(\tau)\hat{\R}^{phy}_{\k_3}(\tau)|C\rangle=\langle 0|\hat{\R}_{\k_1}(\tau)\hat{\R}_{\k_2}(\tau)\hat{\R}_{\k_3}(\tau)|0\rangle.
\end{equation}
So, it is not possible to differentiate between the vacuum state and the coherent states (with constraint (\ref{concoherent})) from the power spectrum or the bispectrum.\\
To understand the result (\ref{nongausscoh}), we can try to rederive it using a simple argument. Just like the two-point function case, we can decompose the field $\hat{\R}^{phy;coh}_{\k}(\tau)$ in the coherent state into $\hat{\R}^{phy;coh}_{\k}(\tau)= \R^{phy;cl}_{\k}(\tau)+\hat{\R}^{vac}_{\k}(\tau)$, with $\R^{phy;cl}_{\k}(\tau)=\langle\hat{\R}^{phy;coh}_{\k}(\tau)\rangle$. By construction $\langle\hat{\R}^{phy;coh}_{\k}(\tau)\rangle$ vanishes in the limit $\tau\rightarrow 0$ and as a result, in this limit $\R^{phy;cl}_{\k}(\tau)$ does not contribute. Therefore, we have $\langle \hat{\R}^{phy;coh}_{\k_1}(\tau)\hat{\R}^{phy;coh}_{\k_2}(\tau)\hat{\R}^{phy;coh}_{\k_3}(\tau)\rangle=\langle \hat{\R}^{vac}_{\k_1}(\tau)\hat{\R}^{vac}_{\k_2}(\tau)\hat{\R}^{vac}_{\k_3}(\tau)\rangle$. However, unlike the two-point function case, here it is not at all obvious that in the presence of interactions the field $\hat{\R}^{phy;coh}_{\k}(\tau)$ can be written as  $\hat{\R}^{phy;coh}_{\k}(\tau)=\hat{\R}^{vac}_{\k}(\tau)+ \R^{phy;cl}_{\k}(\tau)$.


\section{General state}\label{genstate}
Now as the next step, we will compute the power-spectrum with a general initial state and try to find out constraints on the initial state from observations. It is important to note that we are in the Heisenberg picture where states are time-independent. We can write down the most general state using equation (\ref{excitedstate})
\begin{equation}\label{generalstate}
|G\rangle=\sum_{\psi}C_{\psi}|\psi\rangle. 
\end{equation}
And with this initial state we can compute the power spectrum $\langle\hat{\R}_{\k}(\tau)\hat{\R}_{\k'}(\tau)\rangle=\frac{1}{z^2}\langle\hat{v}_{\k}(\tau)\hat{v}_{\k'}(\tau)\rangle$ using
\begin{equation}
\langle\hat{v}_{\k}(\tau)\hat{v}_{\k'}(\tau)\rangle=\frac{\langle G|\hat{v}_{\k}(\tau)\hat{v}_{\k'}(\tau)|G\rangle}{\langle G|G\rangle}.  
\end{equation}
Therefore,
\begin{eqnarray}
\langle\hat{v}_{\k}(\tau)\hat{v}_{\k'}(\tau)\rangle=\frac{1}{2}(2 \pi)^3\delta^3(\k+\k')|u_k(\tau)|^2 + A(\k, \k')u^*_k u^*_{k'} + A^*(-\k, -\k')u_k u_{k'}\nonumber \\ 
+ B(-\k, \k')u_k u^*_{k'}+B(-\k', \k)u^*_k u_{k'}, 
\end{eqnarray}
where,
\begin{equation}
A(\k, \k')=\frac{1}{2}\frac{\langle G|\hat{a}_{\k}\hat{a}_{\k'}|G\rangle}{\langle G|G\rangle}, \qquad B(\k, \k')=\frac{1}{2}\frac{\langle G|\hat{a}^\dagger_{\k}\hat{a}_{\k'}|G\rangle}{\langle G|G\rangle}.
\end{equation}
Introducing $k_*=\sqrt{k k'}$, $\bar{k}=k+k'$ and $\Delta k=k-k'$ and using equation (\ref{modesBD}), we can write
\begin{eqnarray}
\langle\hat{v}_{\k}(\tau)\hat{v}_{\k'}(\tau)\rangle=\frac{1}{2}(2\pi)^3\delta^3(\k+\k')\frac{1}{k}\left(1+\frac{1}{k^2 \tau^2}\right)+A(\k, \k')e^{-i \bar{k} \tau} \frac{1}{k_*}\left(1-\frac{i \bar{k}}{\tau k_*^2}-\frac{1}{k_*^2 \tau^2}\right)\nonumber \\
+A^*(-\k, -\k')e^{i \bar{k} \tau} \frac{1}{k_*}\left(1+\frac{i \bar{k}}{\tau k_*^2}-\frac{1}{k_*^2 \tau^2}\right)+B(-\k, \k')e^{i \Delta k \tau} \frac{1}{k_*}\left(1-\frac{i \Delta k}{\tau k_*^2}+\frac{1}{k_*^2 \tau^2}\right)\nonumber \\
+B(-\k', \k)e^{-i \Delta k \tau} \frac{1}{k_*}\left(1+\frac{i \Delta k}{\tau k_*^2}+\frac{1}{k_*^2 \tau^2}\right).
\end{eqnarray}
In the superhorizon limit ($|k \tau|,|k' \tau|\ll 1$), we have,
\begin{align}\label{finalpowerspec}
\langle\hat{v}_{\k}(\tau)\hat{v}_{\k'}(\tau)\rangle\approx&\frac{1}{2}(2 \pi)^3\delta^3(\k+\k')\frac{1}{k}\left(1+\frac{1}{k^2 \tau^2}\right)\nonumber \\ 
+&\left[-A(\k, \k')-A^*(-\k, -\k')+B(-\k, \k')+B(-\k', \k)\right] 
\left(\frac{1}{k_*^3 \tau^2}+ \frac{k^2+k'^2}{2 k_*^3}\right)+\cdots 
\end{align}
where the dots indicate terms of higher order. Therefore, in the superhorizon limit, at the leading order we have
\begin{align}\label{superhorizonps}
\langle\hat{\R}_{\k}(\tau)\hat{\R}_{\k'}(\tau)\rangle\approx\frac{1}{2} (2 \pi)^3 \frac{ H^4}{ \dot{\bar{\phi}}^2 k^3}\delta^3(\k+\k')
+\frac{ H^4}{ \dot{\bar{\phi}}^2 k_*^3}[&-A(\k, \k')-A^*(-\k, -\k') \nonumber\\
 &+B(-\k, \k')+B(-\k', \k)]. 
\end{align} 

\subsection{Constraints from observations}
In this section, we will try to find the constraints on the initial state from observations. Our universe as we see it today, is homogeneous and isotropic on large scale. Temperature fluctuations of CMB and LSS are directly related to the curvature perturbations produced during inflation. Also the present observations of the CMB temperature inhomogeneities indicates the presence of almost scale-invariant spectrum of curvature perturbations. In general, the one-point function $\langle\hat{\R}_{\k}(\tau)\rangle \neq 0$, even though $\hat{\R}_{\k}(\tau)$ is a perturbation.\footnote{It is easy to understand this by considering the simpler case $S=\int dt\left[\frac{1}{2}\dot{x}^2-V(x)\right] $. Action for the perturbation around the classical solution $\bar{x}$ is given by, $S=\int dt\left[\frac{1}{2}\dot{\delta x}^2-\frac{1}{2}V''(\bar{x})(\delta x)^2\right] $. This perturbations can be quantized just like harmonic oscillator and thus in general $\langle \Psi|\hat{\delta x}(t)|\Psi\rangle\neq 0$ for a general quantum state $|\Psi\rangle$.} Homogeneity demands that $\langle\hat{\R}_{\k}(\tau)\rangle=0$ but observation requires homogeneity only for the superhorizon modes. Therefore we will impose $\langle\hat{\R}_{\k}(\tau)\rangle=0$ only in the superhorizon limit. From that we can write down the first constraint  
\begin{equation}\label{firstcon}
\langle G|\hat{a}^\dagger_{-\k}|G\rangle=\langle G|\hat{a}_{\k}|G\rangle. 
\end{equation}
Homogeneity also demands that, in the superhorizon limit
\begin{equation}\label{twopoint}
\langle\hat{\R}_{\k}(\tau)\hat{\R}_{\k'}(\tau)\rangle= P(\k, \tau)\delta^3(\k+\k'). 
\end{equation}
Where $P(\k, \tau)$ is some arbitrary function of $\k$ and $ \tau$. And the isotropy condition tells us that $P(\k, \tau)=P(k,\tau)$. Now demanding scale-invariance we get 
\begin{equation}\label{genP}
P(k, \tau)=(2 \pi)^3\frac{ H^4}{ \dot{\bar{\phi}}^2}\frac{W}{k^3}, 
\end{equation}
where $W$ is some dimensionless constant independent of $k$. Factors $(2 \pi)^3$ and $ H^4/ \dot{\bar{\phi}}^2$ are included for later convenience. Comparing the last equation with the leading order term of equation (\ref{superhorizonps}), we also find that $W$ does not depend on $\tau $ and hence $\langle\hat{\R}_{\k}(\tau)\hat{\R}_{\k'}(\tau)\rangle$ is time-independent. Now comparing equation (\ref{twopoint}) with the equation (\ref{superhorizonps}), we get the second constraint equation
\begin{equation}\label{secondcon}
 -A(\k, \k')-A^*(-\k, -\k')+B(-\k, \k')+B(-\k', \k)=(2 \pi)^3 W' \delta^3(\k+\k'),  
\end{equation}
where $W'=(W-\frac{1}{2})$ is also a constant. And the power spectrum is given by
\begin{equation} 
\langle\hat{\R}_{\k}(\tau)\hat{\R}_{\k'}(\tau)\rangle= (2 \pi)^3 \frac{ H^4}{ \dot{\bar{\phi}}^2 k^3}\left(\frac{1}{2}+W'\right)\delta^3(\k+\k'),
\end{equation}
where $W'$ is defined by equation (\ref{secondcon}). Equations (\ref{firstcon}, \ref{secondcon}) are two constraints on the initial state.\footnote{It is possible to avoid the first constraint (\ref{firstcon}) by introducing a new field $\hat{\R}_{\k}^{new}(\tau)=\hat{\R}_{\k}(\tau)-\langle\hat{\R}_{\k}(\tau)\rangle$ and defining the power spectrum as $\langle\hat{\R}_{\k}^{new}(\tau)\hat{\R}_{\k'}^{new}(\tau)\rangle$. That will obviously modify the second constraint (\ref{secondcon}).} It is reasonable to assume that the initial state $|G\rangle$ does not contain any excitation of infinitely large momentum. That is $W'=0$. In that case constraint (\ref{secondcon}) reads
\begin{equation}\label{thirdcon}
 -A(\k, \k')-A^*(-\k, -\k')+B(-\k, \k')+B(-\k', \k)=0  
\end{equation}
with power spectrum
\begin{equation} 
\langle\hat{\R}_{\k}(\tau)\hat{\R}_{\k'}(\tau)\rangle=\frac{1}{2} (2 \pi)^3 \frac{ H^4}{ \dot{\bar{\phi}}^2 k^3}\delta^3(\k+\k').
\end{equation}

\subsection{Some examples}
Now we will give some examples to show that it is possible to construct states that satisfy constraints (\ref{firstcon}) and (\ref{thirdcon}). Vacuum state $|G\rangle=|0\rangle$ is obviously the most trivial example of a state that satisfies both these constraints. It is also easy to check that coherent states with condition (\ref{concoherent}) satisfy both constraints (\ref{firstcon}) and (\ref{thirdcon}).

To give another nontrivial example we look for states of the form $|G\rangle=|0\rangle + |\psi\rangle$, where $|\psi\rangle$ is an excited state (or a combination of excited states). The simplest example of such a state is $|G\rangle=|0\rangle + \int d^3\k ~ \alpha(k)~\hat{a}^\dagger_{\k} |0\rangle$. But it is easy to check that this state does not work. $\alpha(k)$ has to be real to satisfy equation (\ref{firstcon}). Then one can show that $\alpha(k)$ has to vanish to satisfy equation (\ref{thirdcon}). But  
\begin{equation}\label{exone}
|G\rangle=|0\rangle + \int d^3\k ~ \alpha(k)~\hat{a}^\dagger_{\k} |0\rangle +\int d^3\k_1 d^3\k_2 ~ \beta(k_1)\beta(k_2)~\hat{a}^\dagger_{\k_1} \hat{a}^\dagger_{\k_2}|0\rangle,  
\end{equation}
where $\alpha(k)$ and $\beta(k)$ are real functions, makes it possible to construct states that satisfy both constraints (\ref{firstcon}) and (\ref{thirdcon}) and hence produce a scale-invariant power spectrum. Equation (\ref{firstcon}) is already satisfied and equation (\ref{thirdcon}) leads to 
\begin{equation}\label{conexone}
\alpha(k)\alpha(k')+ (4N-1)\beta(k)\beta(k')=0 
\end{equation}
with $N=(2 \pi)^3 \int d^3\k \left(\beta(k)\right)^2$. So all the states given by equation (\ref{exone}) with $\alpha(k)$ and $\beta(k)$ obeying equation (\ref{conexone}) produce scale invariant power spectrum. Example of one such state is
\begin{equation}\label{extwo}
|G\rangle=|0\rangle + A \sqrt{1-\frac{32 \pi^4 A^2}{\gamma^3}}\int d^3\k ~ e^{-\gamma k}~\hat{a}^\dagger_{\k} |0\rangle +A^2\int d^3\k_1 d^3\k_2 ~ e^{-\gamma(k_1+k_2)}~\hat{a}^\dagger_{\k_1} \hat{a}^\dagger_{\k_2}|0\rangle,  
\end{equation}
where, $A$ and $\gamma$ are real constants and $\frac{32 \pi^4 A^2}{\gamma^3}\leq 1$. Renormalizability of the energy-momentum tensor of the fluctuations can impose more constraints on these states. Here we should note that it is possible to construct infinite number of such states (with even more complicated combination of excited states) that obey constraints (\ref{firstcon}) and (\ref{thirdcon}) and hence can produce a scale-invariant power-spectrum.


\section{Constraints from renormalizability of the energy-momentum tensor}\label{renormalization}
\subsection{Energy momentum tensor for perturbations}
In the longitudinal gauge, the metric has the form
\begin{equation}
ds^2=-(1+2\Phi)dt^2+a^2(t)\left[(1-2\Psi)\delta_{ij}\right]dx^idx^j.
\end{equation}
The Einstein's equation is $G_{\mu\nu}-8 \pi G T_{\mu\nu}\equiv \Pi_{\mu\nu}=0$. To lowest order background obeys $\Pi^{(0)}_{\mu\nu}=0$. The first order Einstein's equations $\Pi^{(1)}_{\mu\nu}=0$ give the equations of motion for the perturbations. At first order we get $\Phi=\Psi$. Other equations of motion for the perturbations are
\begin{align}
\dot{\Psi}+H\Psi=&4 \pi G \dot{\bar{\phi}} \delta\phi, \label{firsteq} \\
\delta\ddot{\phi}+3 H \delta\dot{\phi}+ V''(\bar{\phi}) \delta\phi-\left(\frac{\nabla^2}{a^2}\right) \delta\phi=&-2\Psi V'(\bar{\phi})+4\dot{\Psi}\dot{\bar{\phi}}, \\
-\left(4 \pi G \dot{\bar{\phi}}^2 + \frac{\nabla^2}{a^2}\right)\Psi=& 4 \pi G\left(- \dot{\bar{\phi}} \delta\dot{\phi}+\ddot{\bar{\phi}}\delta\phi\right).
\end{align}
We have the good old gauge invariant variable $\R$ defined by,
\begin{equation}
\R= \Psi + \frac{H}{\dot{\bar{\phi}}} \delta \phi.
\end{equation}
Following \cite{Abramo:1997hu}, we can write down the energy-momentum tensor for the perturbations $t_{\mu\nu}=-\Pi^{(2)}_{\mu\nu}$ in terms of $\Psi$ and $\delta\phi$
\begin{align}\label{emtensor}
t_{00}=\frac{1}{8 \pi G}&\left[12H(\Psi\dot{\Psi}) - 3(\dot{\Psi})^2+ 9a^{-2}(\nabla \Psi)^2\right]\nonumber\\
&+\left[\frac{1}{2}(\delta\dot{\phi})^2+ \frac{1}{2}a^{-2}(\nabla \delta\phi)^2+\frac{1}{2}V''(\bar{\phi})(\delta\phi)^2+ 2 V'(\bar{\phi})(\Psi \delta\phi)\right],\\
t_{ij}=a^2 \delta_{ij}&\left(\frac{1}{8 \pi G}\left[(24H^2+16\dot{H})\Psi^2+24H(\Psi\dot{\Psi}) +(\dot{\Psi})^2+4\Psi \ddot{\Psi}-\frac{4}{3}a^{-2}(\nabla \Psi)^2\right]\right.\nonumber \\
&\left.+\left[4\dot{\bar{\phi}}^2\Psi^2+\frac{1}{2} (\delta\dot{\phi})^2- \frac{1}{6}a^{-2}(\nabla \delta\phi)^2-4\dot{\bar{\phi}}(\delta\dot{\phi}\Psi)-\frac{1}{2}V''(\bar{\phi})(\delta\phi)^2+ 2 V'(\bar{\phi})(\Psi \delta\phi)\right]\right).
\end{align}
Our goal is to find out constraints on the initial state from renormalizability of the energy-momentum tensor. So we want to consider the contributions of large-$k$ fluctuations of $t_{\mu\nu}$. At large-$k$ $(k>>aH)$,
\begin{equation}
\R_k\sim \frac{H}{a \dot{\bar{\phi}}} \frac{e^{i k \tau}}{\sqrt{k}}.
\end{equation}
From equation (\ref{firsteq}) we have for large-$k$ (and for quasi-de Sitter inflation)
\begin{equation}
\Psi_k\sim -\frac{i H \epsilon a}{k}\R_k,
\end{equation}
where, $\epsilon$ is the slow roll parameter. In this approximation, for $\delta\phi$ we obtain,
\begin{equation}\label{delphi}
\delta\phi\sim\R \frac{\dot{\bar{\phi}}}{H}\left(1+\frac{i H \epsilon a}{k}\right)\sim\R \frac{\dot{\bar{\phi}}}{H}.
\end{equation}
By inspection, it is very clear that all terms in equation (\ref{emtensor}) containing $\Psi$ are suppressed by powers of $Ha/k$ compared to terms without $\Psi$. Therefore in large-$k$ limit we have,
\begin{align}
t_{00}\approx&\left[\frac{1}{2}(\delta\dot{\phi})^2+ \frac{1}{2}a^{-2}(\nabla \delta\phi)^2+\frac{1}{2}V''(\bar{\phi})(\delta\phi)^2\right],\\
t_{ij}\approx& a^2 \delta_{ij}\left[\frac{1}{2} (\delta\dot{\phi})^2- \frac{1}{6}a^{-2}(\nabla \delta\phi)^2\frac{1}{2}V''(\bar{\phi})(\delta\phi)^2\right].
\end{align}


\subsection{Renormalization of energy-momentum tensor}
Using equation (\ref{delphi}), we can write down $t_{\mu \nu}$ (for large-$k$) in terms of gauge invariant variable $\R$
\begin{align}
\langle \hat{t}_{00}\rangle \approx&\left(\frac{\dot{\bar{\phi}}}{H}\right)^2\left[\frac{1}{2}a^{-2}\langle(\hat{\R}')^2\rangle+ \frac{1}{2}a^{-2}\langle(\nabla \hat{\R})^2\rangle+\frac{1}{2}V''(\bar{\phi})\langle\hat{\R}^2\rangle\right],\\
\langle \hat{t}_{ij}\rangle \approx&a^2 \delta_{ij}\left(\frac{\dot{\bar{\phi}}}{H}\right)^2\left[\frac{1}{2}a^{-2}\langle(\hat{\R}')^2\rangle- \frac{1}{6}a^{-2}\langle(\nabla \hat{\R})^2\rangle-\frac{1}{2}V''(\bar{\phi})\langle\hat{\R}^2\rangle\right].
\end{align}
For the vacuum state, we have the following expressions for the unregularized energy-momentum tensor
\begin{align}
\langle 0| \hat{t}_{00}|0 \rangle \approx&\frac{1}{4 \pi^2}\int^{\infty} dk H^4\left[\frac{3 \eta}{2k}+ \frac{1}{2}\tau^2 k (1+3\eta) +\tau^4 k^3\right],\\
\langle 0| \hat{t}_{ij}|0\rangle \approx& -a^2 \delta_{ij}\frac{1}{4 \pi^2}\int^{\infty} dk H^4\left[\frac{3 \eta}{2k}+ \frac{1}{6} \tau^2 k (1+9\eta) -\frac{1}{3}\tau^4 k^3\right],
\end{align}
where $\eta$ is the second slow-roll parameter. To obtain the renormalized value of $\langle \hat{t}_{\mu \nu}\rangle$, one can use any regularization method (for example adiabatic regularization).  Detailed discussions of adiabatic regularization method can be found in \cite{Parker:1974qw, Bunch:1980vc}. Adiabatic regularization of $\langle \hat{t}_{\mu \nu}\rangle$ can be done by subtracting adiabatic modes up to order four.
\begin{equation}\label{rennemtensor}
\langle 0|  \hat{t}_{\mu \nu}|0 \rangle_{ren}=\langle 0|  \hat{t}_{\mu \nu}|0 \rangle-\langle 0|  \hat{t}_{\mu \nu}|0 \rangle_{adi},
\end{equation} 
where, $\langle 0|  \hat{t}_{\mu \nu}|0 \rangle_{adi}$ is calculated using the adiabatic mode functions of order four
\begin{equation}
u_k^{adi;4}(\tau)= \frac{1}{\sqrt{W^{(4)}}} e^{i \int^{\tau} W^{(4)}d \tau}, \qquad \R^{adi}_k(\tau)= \left(\frac{H}{a \dot{\bar{\phi}}}\right)u_k^{adi;4}(\tau),
\end{equation}
where,
\begin{equation}
W^{(4)}=k-\frac{1}{k \tau^2}+\frac{1}{k^3 \tau^4}.
\end{equation}
Therefore we have the following expression for  $\langle 0|  \hat{t}_{\mu \nu}|0 \rangle_{adi}$
\begin{align}
\langle 0| \hat{t}_{00}|0 \rangle_{adi} \approx&\frac{1}{4 \pi^2}\int^{\infty} dk H^4\left[\frac{3 \eta}{2k}+ \frac{1}{2}\tau^2 k (1+3\eta) +\tau^4 k^3\right] +\text{UV finite},\\
\langle 0| \hat{t}_{ij}|0\rangle_{adi} \approx& -a^2 \delta_{ij}\frac{1}{4 \pi^2}\int^{\infty} dk H^4\left[\frac{3 \eta}{2k}+ \frac{1}{6} \tau^2 k (1+9\eta) -\frac{1}{3}\tau^4 k^3\right]+\text{UV finite}.
\end{align}
Hence, $\langle 0|  \hat{t}_{\mu \nu}|0 \rangle_{ren}$ does not have any ultraviolet divergences. For a general initial state $|G\rangle$ we can renormalize energy-momentum tensor by
\begin{equation}\label{rengenstate}
\langle G|  \hat{t}_{\mu \nu}|G \rangle_{ren}\equiv\langle G|  \hat{t}_{\mu \nu}|G \rangle-\langle 0|  \hat{t}_{\mu \nu}|0 \rangle_{adi}.
\end{equation} 
Note that in the right hand side, we do not have $\langle G|  \hat{t}_{\mu \nu}|G \rangle_{adi}$. But we have $\langle 0|  \hat{t}_{\mu \nu}|0 \rangle_{adi}$. It can be understood easily in Minkowski space limit. In Minkowski space, $\langle G|  \hat{t}_{\mu \nu}|G \rangle_{adi}=\langle G|  \hat{t}_{\mu \nu}|G \rangle $. So, if we had $\langle G|  \hat{t}_{\mu \nu}|G \rangle_{adi}$ in equation (\ref{rengenstate}), then $\langle G|  \hat{t}_{\mu \nu}|G \rangle_{ren}=0$ for any state $|G\rangle$. Clearly that can not be right. \\

We will impose the constraint on the initial state $|G\rangle$ that $\langle G|  \hat{t}_{\mu \nu}|G \rangle_{ren}$ does not have any ultraviolet divergences. That means the state $|G\rangle$ does not introduce any new ultra-violet divergences to the energy-momentum tensor. Therefore, to make sure that $\langle G|  \hat{t}_{\mu \nu}|G \rangle_{ren}$ has desired UV behavior, we will impose the following constraint on the initial state
\begin{equation}\label{emconstraint}
\langle G|  \hat{t}_{\mu \nu}|G \rangle=\langle 0|  \hat{t}_{\mu \nu}|0 \rangle+\text{UV finite}.
\end{equation}
If we impose constraint (\ref{emconstraint}) on the coherent state $\hat{a}_{\k}|C\rangle=C(\k)|C\rangle$ (with $C^{*}(-\k)=C(\k)$), we find that $C(\k)$ goes to zero faster than $\frac{1}{k^{5/2}}$ for large $\k$. One can check that state (\ref{extwo}) also satisfies constraint (\ref{emconstraint}).


\section{Bispectrum in the squeezed limit}\label{bispectrum}
In principle one can calculate the three-point function in the momentum space using an expression similar to equation (\ref{effthreepointfun}). At leading order in slow-roll parameter, just like the coherent state case, we can write
\begin{align}\label{generalfnl}
\langle G|\hat{\R}^{phy}_{\k_1}(\tau)\hat{\R}^{phy}_{\k_2}(\tau)\hat{\R}^{phy}_{\k_3}(\tau)|G\rangle=&\langle G|\hat{\R}_{\k_1}(\tau)\hat{\R}_{\k_2}(\tau)\hat{\R}_{\k_3}(\tau)|G\rangle \nonumber \\
&-\left(\langle G|\hat{\R}_{\k_1}(\tau)|G\rangle \langle G|\hat{\R}_{\k_2}(\tau)\hat{\R}_{\k_3}(\tau)|G\rangle+ \text{cyclic perm}\right).
\end{align}
In practice, it is not very easy to compute three-point function for any general state $|G\rangle$. Creminelli and Zaldarriaga \cite{Creminelli:2004yq} (see also \cite{Ganc:2010ff} for a good discussion) used a clever argument to determine the three-point function in the local limit (i.e. $k_1, k_2\gg k_3$). In the local limit, bispectrum has the form
\begin{equation}
\langle\hat{\R}_{\k_1}(\tau)\hat{\R}_{\k_2}(\tau)\hat{\R}_{\k_3}(\tau)\rangle_{k_3\ll k_1\approx k_2}=(2 \pi)^3 \delta^3(\k_1+\k_2+\k_3)\frac{12}{5}f_{NL}^{loc}P_R(k_3)P_R(k_1).
\end{equation}
They found out for any single field inflation model
\begin{equation}
f_{NL}^{loc}=\frac{5}{12}(1-n_s) + O\left(\frac{k_3}{k_1}\right)^2.
\end{equation}
It is possible to extend their argument even for any general initial state. We want to compute $\langle G|\hat{\R}_{\k_1}(\tau)\hat{\R}_{\k_2}(\tau)\hat{\R}_{\k_3}(\tau)|G\rangle$ after $k_1, k_2$ modes have crossed the horizon. $k_3$ mode will have crossed the horizon in the distant past because $k_1, k_2\gg k_3$. Therefore, $k_3$ mode behaves classically (see \cite{Creminelli:2004yq}) and contributes to the background metric\footnote{Now we are working in the comoving gauge.}
\begin{equation}\label{metriclocal}
ds^2=-dt^2+a^2(t)e^{-2\R_B(\mathbf{x})}d\mathbf{x}^2,
\end{equation}
where, $\R_B$ is the contributions from modes far outside the horizon
\begin{equation}
\R_B(\mathbf{x}, \tau)=\int_{k\ll k_1, k_2} \frac{d^3\mathbf{k}}{(2\pi)^{3}}  \R_{\mathbf{k}}e^{i \k.\mathbf{x}}.
\end{equation}
Therefore, in the local limit ($k_3\ll k_1\approx k_2$) we have,
\begin{equation}
\langle G|\hat{\R}_{\k_1}(\tau)\hat{\R}_{\k_2}(\tau)\hat{\R}_{\k_3}(\tau)|G\rangle\approx\langle G|\langle\hat{\R}_{\k_1}(\tau)\hat{\R}_{\k_2}(\tau)\rangle_{\R_{\k_3}}\hat{\R}_{\k_3}(\tau)|G\rangle.
\end{equation}
Where, $\langle\hat{\R}_{\k_1}(\tau)\hat{\R}_{\k_2}(\tau)\rangle_{\R_{\k_3}}=\langle G|\hat{\R}_{\k_1}(\tau)\hat{\R}_{\k_2}(\tau)|G\rangle_{\R_{\k_3}}$ is the two-point function in the perturbed background (\ref{metriclocal}). It is clear from the metric (\ref{metriclocal}) that long wavelength mode (neglecting the gradients) is equivalent to the rescaling of the coordinates
\begin{equation}\label{rescaling}
\mathbf{x} \rightarrow \mathbf{x}'=\Lambda \mathbf{x},
\end{equation}
where, $\Lambda= e^{-\R_B}$. Under this rescaling of the coordinates, it is easy to check that $\R_{\k}$ transforms as
\begin{equation}
\R_{\k} \rightarrow \Lambda^3\R_{\k \Lambda}=\Lambda^3\R_{\bar{\k}}.
\end{equation}
Therefore, 
\begin{align}
\langle G|\hat{\R}_{\k_1}(\tau)\hat{\R}_{\k_2}(\tau)|G\rangle_{\R_{\k_3}}&=\Lambda^6 \langle G|\hat{\R}_{\bar{\k}_1}(\tau)\hat{\R}_{\bar{\k}_2}(\tau)|G\rangle \nonumber \\
&=\Lambda^6(2 \pi)^3 P_{\R}(\bar{k}_1)\delta^3(\bar{\k}_1+\bar{\k}_2) \nonumber \\
&=\Lambda^6(2 \pi)^3 P_{\R}(k_1)(\Lambda^{-4+n_s})\delta^3(\k_1+\k_2)\frac{1}{\Lambda^3} \nonumber\\
&=(2 \pi)^3 P_{\R}(k_1)(\Lambda^{-1+n_s})\delta^3(\k_1+\k_2) \nonumber\\
&=(2 \pi)^3 P_{\R}(k_1)\left(e^{\R_B(1-n_s)}\right)\delta^3(\k_1+\k_2) \nonumber\\
&=(2 \pi)^3 P_{\R}(k_1)\left[1+ \R_B(1-n_s)+...\right]\delta^3(\k_1+\k_2). 
\end{align}
Only thing we have assumed here is that in the superhorizon limit, $\langle G|\hat{\R}_{\k_1}(\tau)\hat{\R}_{\k_2}(\tau)|G\rangle\propto \frac{1}{k^{(4-n_s)}}\delta^3(\k_1+\k_2)$. So far we have treated $\R_{\k_3}$ as a classical field. Now we will promote $\R_{\k_3}$ to an quantum operator. That leads to
\begin{align}
\langle G|\hat{\R}_{\k_1}(\tau)&\hat{\R}_{\k_2}(\tau)\hat{\R}_{\k_3}(\tau)|G\rangle  \approx\langle G|\langle\hat{\R}_{\k_1}(\tau)\hat{\R}_{\k_2}(\tau)\rangle_{\R_{\k_3}}\hat{\R}_{\k_3}(\tau)|G\rangle \nonumber \\
&= (2 \pi)^3 P_{\R}(k_1)\delta^3(\k_1+\k_2)(1-n_s) \int_{k\ll k_1, k_2} \frac{d^3\mathbf{k}}{(2\pi)^{3}}e^{i \k.\mathbf{x}}\langle G|\hat{\R}_{\k}(\tau)\hat{\R}_{\k_3}(\tau)|G\rangle \nonumber \\
& \approx (2 \pi)^3 P_{\R}(k_1)P_{\R}(k_3)\delta^3(\k_1+\k_2)(1-n_s) \int_{k\ll k_1, k_2}d^3\mathbf{k}e^{i \k.\mathbf{x}}\delta^3(\k+\k_3)\nonumber \\
& \approx (2 \pi)^3 P_{\R}(k_1)P_{\R}(k_3)(1-n_s)\delta^3(\k_1+\k_2).
\end{align}
Therefore, in the squeezed limit we have,
\begin{equation}\label{localfnl}
f_{NL}^{loc}\approx\frac{5}{12}(1-n_s). 
\end{equation}
If the state $|G\rangle$ produces almost scale invariant power spectrum then it can not produce large $f_{NL}^{loc}$. This result is actually true for any single field inflation model with a general initial state for primordial fluctuations in the limit $k_3\rightarrow 0$. For finite $k_3$, the correction to  (\ref{localfnl}) can be large for some particular states (as reported in \cite{Agullo:2010ws, Ganc:2011dy}). Therefore for finite $k_3$, one should be more careful and for a given state a detailed calculation should be performed using (\ref{generalfnl}).


\section{Conclusions}\label{conclusions}
In this paper, we have explored the possibility of a general initial state for primordial fluctuations. Constraints on initial state from current measurements of power spectrum and bispectrum are relatively weak and for quasi-de Sitter inflation (or slow roll inflation), a large number of states are consistent with the observations. Vacuum state is just one such example. Coherent states with constraint (\ref{concoherent}) are also interesting examples of states that are consistent with current observations. It is impossible to differentiate between these coherent states and the Bunch-Davies vacuum state just from the power-spectrum and the bispectrum.  Although we have not considered tensor modes here, similar analysis can be done for gravitational waves with similar results.

Most of the results in this paper only apply to canonical, single-field inflationary model with equation of state parameter $w$ approximately constant during inflationary epoch. But the most significant limitation is the ambiguity in the definition of vacuum. We have used Bunch-Davies prescription of vacuum throughout the paper. The only justification for it, is the fact that for the modes deep inside horizon, space looks Minkowski and hence the Bunch-Davies vacuum is a reasonable choice.

It is clear from our discussion (section \ref{genstate} and \ref{bispectrum}) that a large class of general initial states can produce scale invariant power spectrum and small $f_{NL}^{loc}$ just like the Bunch-Davies vacuum state. But higher N-point functions can be a very useful tool for probing the initial state. The PLANCK satellite will significantly increase the precision of CMB observations. Observations made by PLANCK may contain valuable information about the initial state of primordial fluctuations and that would provide a window for the physics before inflation.

\section*{Acknowledgments}
I particularly want to thank Willy Fischler and Eiichiro Komatsu for numerous helpful discussions. I would also like to thank Arnab Kundu, Joel Meyers, Dan Carney, Anindya Dey for discussions. This material is based upon work supported by the National Science Foundation under Grant Number PHY-0969020 and by the Texas Cosmology Center, which is supported by the College of Natural Sciences and the Department of Astronomy at the University of Texas at Austin and the McDonald Observatory.

\appendix

\section{Computation of three-point function with coherent state}
\label{appendix}

Using time-dependent perturbation theory we have,
\begin{align}\label{threepoint}
\langle C|&\hat{\R}_{\k_1}(\tau)\hat{\R}_{\k_2}(\tau)\hat{\R}_{\k_3}(\tau)|C\rangle\nonumber\\
&=\langle C|\left(\bar{T} e^{i \int_{\tau_0}^{\tau}H^{I}_{int}(\tau')d\tau'} \right) \hat{\R}_{\k_1}^{I}(\tau)\hat{\R}_{\k_2}^{I}(\tau)\hat{\R}_{\k_3}^{I}(\tau)\left(T e^{-i \int_{\tau_0}^{\tau}H^{I}_{int}(\tau')d\tau'} \right)|C\rangle \nonumber \\
&= \langle C|\hat{\R}_{\k_1}^{I}(\tau)\hat{\R}_{\k_2}^{I}(\tau)\hat{\R}_{\k_3}^{I}(\tau)|C\rangle-i\int_{\tau_0}^{\tau}d\tau'\langle C|\left[\hat{\R}_{\k_1}^{I}(\tau)\hat{\R}_{\k_2}^{I}(\tau)\hat{\R}_{\k_3}^{I}(\tau), H^{I}_{int}(\tau')\right]|C\rangle.
\end{align}
Now all the fields are in the interaction picture and $H_{int}$ is given by equation (\ref{Hint}). $T$ and $\bar{T}$ are the time and anti-time ordered product respectively. $\tau_0$ is the conformal time at the beginning of inflation and we will take the limit $\tau_0\rightarrow -\infty$.\footnote{For large $\tau_0$, all the exponentials with $\tau_0$ will oscillate. When performing our calculations, we will use the average value (i.e. zero) for them.} Throughout the calculation we will assume that $\k_i\neq0$. We will also take the usual limit $\tau\rightarrow 0$. The first term in (\ref{threepoint}) can be written using the redefined field (\ref{rc})
\begin{align}\label{pert}
\langle C|\hat{\R}_{\k_1}^{I}(\tau)&\hat{\R}_{\k_2}^{I}(\tau)\hat{\R}_{\k_3}^{I}(\tau)|C\rangle=\langle C|\hat{\R}_{c,\k_1}^{I}(\tau)\hat{\R}_{c,\k_2}^{I}(\tau)\hat{\R}_{c,\k_3}^{I}(\tau)|C\rangle &\\
+ &\frac{1}{4}\left(3\epsilon -2\eta\right)\left(\int \frac{d^3\mathbf{\p}}{(2\pi)^{3}}\langle C|\hat{\R}_{c,\k_1}^{I}(\tau)\hat{\R}_{c,\k_2}^{I}(\tau)\hat{\R}_{c,\p}^{I}(\tau)\hat{\R}_{c,\k_3-\p}^{I}(\tau)|C\rangle+\text{cyclic perm}\right) \nonumber \\
+ &\frac{1}{2} \epsilon\left(\int \frac{d^3\mathbf{\p}}{(2\pi)^{3}}\frac{(\k_3-\p)^2}{k_3^2}\langle C|\hat{\R}_{c,\k_1}^{I}(\tau)\hat{\R}_{c,\k_2}^{I}(\tau)\hat{\R}_{c,\p}^{I}(\tau)\hat{\R}_{c,\k_3-\p}^{I}(\tau)|C\rangle+\text{cyclic perm}\right). \nonumber
\end{align}
$\hat{\R}_{c,\k}^{I}(\tau)$ behaves like the free field, and can be written as
\begin{equation}
\hat{\R}_{c,\k}^{I}(\tau)=\frac{1}{\sqrt{2}}\left[\hat{a}_{\k}\R^*_k(\tau)+ \hat{a}^\dagger_{-\k}\R_k(\tau)\right],
\end{equation}
where $\R_k(\tau)=\left(\frac{H}{a \dot{\bar{\phi}}} \right) \frac{e^{ik\tau}}{\sqrt{k}}\left(1+\frac{i}{k\tau}\right)$. Using the commutation relation
\begin{equation}
\left[\hat{\R}_{c,\k}^{I}(\tau), \hat{a}^\dagger_{\k'}\right]=\frac{1}{\sqrt{2}}(2 \pi)^3 \R^*_k(\tau)\delta^3(\k-\k'),
\end{equation}
the first term in the right hand side of equation (\ref{pert}) can be calculated
\begin{align}
\langle C|\hat{\R}_{c,\k_1}^{I}(\tau)\hat{\R}_{c,\k_2}^{I}(\tau)\hat{\R}_{c,\k_3}^{I}(\tau)|C\rangle=\frac{1}{\sqrt{2}}
(2 \pi)^3 C(\k_3)]2Re[\R_{k_3}(\tau)] )P_R(k_1)\delta^3(\k_1+\k_2)& \nonumber\\
+ \left[\frac{1}{2}(2 \pi)^3 \R_{k_2}(\tau)\R^*_{k_2}(\tau)\langle C|\hat{\R}_{c,\k_1}^{I}(\tau)|C\rangle\delta^3(\k_3+\k_2)+ \k_1\leftrightarrow\k_2 \right].& 
\end{align} 
In the limit $\tau\rightarrow 0$, $\langle C|\hat{\R}_{c,\k}^{I}(\tau)|C\rangle=0$ because of the constraint (\ref{concoherent}) and $\R_{k}(\tau)$ is purely imaginary. Therefore,
\begin{equation}
\langle C|\hat{\R}_{c,\k_1}^{I}(\tau)\hat{\R}_{c,\k_2}^{I}(\tau)\hat{\R}_{c,\k_3}^{I}(\tau)|C\rangle=0.
\end{equation}
Last two terms can also be computed and in the limit $\tau\rightarrow 0$ we get,
\begin{align}
\langle C|\hat{\R}_{\k_1}^{I}(\tau)\hat{\R}_{\k_2}^{I}(\tau)&\hat{\R}_{\k_3}^{I}(\tau)|C\rangle= (2 \pi)^3 \left[\frac{1}{2}\left(3\epsilon -2\eta\right)\left(P_R(k_2)P_R(k_1) + \text{cyclic perm}\right)\right.\nonumber \\ 
+&\left.\frac{1}{2} \epsilon\left(P_R(k_2)P_R(k_1)\frac{k_1^2+k_2^2}{k_3^2} + \text{cyclic perm}\right)\right]\delta^3(\k_1+\k_2+\k_3).
\end{align}
Next we will compute,
\begin{align}
&\int_{\tau_0}^{\tau}d\tau'\langle  C|\hat{\R}_{\k_1}^{I}(\tau)\hat{\R}_{\k_2}^{I}(\tau)\hat{\R}_{\k_3}^{I}(\tau) H^{I}_{int}(\tau')|C\rangle=2 \epsilon \int_{\tau_0}^{\tau}d\tau' a^3(\tau')\left( \frac{\dot{\bar{\phi}}^2}{H}\right)\int d^3\p_1 d^3\p_2 d^3\p_3 \delta^3(\p) \nonumber \\
& \times \left(\frac{1}{p_3^2}\right)[(Re[\R'_{p_3}(\tau')]Re[\R'_{p_2}(\tau')]C(\p_3)C(\p_2)\R^*_{k_3}(\tau)\R'_{k_3}(\tau')P_R(k_1)\delta^3(\k_1+\k_2)\delta^3(\k_3+\p_1) \nonumber \\
&+ \text{p-cyclic})+\text{k-cyclic}] -2\epsilon \frac{(2 \pi)^3}{i (k_1+k_2+k_3)}\delta^3(\k)\left[P_R(k_1)P_R(k_2)\frac{k_1^2k_2^2}{k_3^3} + \text{cyclic perm}\right].
\end{align}
Where, $\p=\p_1+\p_2+\p_3$, $\k=\k_1+\k_2+\k_3$ and we have used the following equation
\begin{equation}
\R^*_{k}(\tau)\R'_{k}(\tau')=-\left( \frac{H^3}{\dot{\bar{\phi}}^2}\right)\frac{1}{ a(\tau') k}e^{ik\tau'}.
\end{equation}
Similarly,
\begin{align}
&\int_{\tau_0}^{\tau}d\tau'\langle  C| H^{I}_{int}(\tau')\hat{\R}_{\k_1}^{I}(\tau)\hat{\R}_{\k_2}^{I}(\tau)\hat{\R}_{\k_3}^{I}(\tau)|C\rangle= 2 \epsilon \int_{\tau_0}^{\tau}d\tau' a^3(\tau')\left( \frac{\dot{\bar{\phi}}^2}{H}\right)\int d^3\p_1 d^3\p_2 d^3\p_3 \delta^3(\p) \nonumber \\
& \times \left(\frac{1}{p_3^2}\right) [(Re[\R'_{p_3}(\tau')]Re[\R'_{p_2}(\tau')]C(\p_3)C(\p_2)\R_{k_3}(\tau)\R^*{'}_{k_3}(\tau')P_R(k_1)\delta^3(\k_1+\k_2)\delta^3(\k_3+\p_1) \nonumber \\
&+ \text{p-cyclic})+\text{k-cyclic}]+2\epsilon \frac{(2 \pi)^3}{i (k_1+k_2+k_3)}\delta^3(\k)\left[P_R(k_1)P_R(k_2)\frac{k_1^2k_2^2}{k_3^3} + \text{cyclic perm}\right].
\end{align}
Now we have to compute the last term in equation (\ref{effthreepointfun})
\begin{align}
\langle C |\hat{\R}_{\k_1}(\tau)|C\rangle \langle C|\hat{\R}_{\k_2}(\tau)\hat{\R}_{\k_3}(\tau)|C\rangle=2 \epsilon \int_{\tau_0}^{\tau}d\tau' a^3(\tau')\left( \frac{\dot{\bar{\phi}}^2}{H}\right)&\int  d^3\p_1 d^3\p_2 d^3\p_3 \delta^3(\p)\left(\frac{1}{p_3^2}\right)  \nonumber \\
\times (2Re[\R'_{p_3}(\tau')]Re[\R'_{p_2}(\tau')]C(\p_3)C(\p_2)Im[\R^*_{k_1}(\tau)\R'_{k_1}(\tau')]& P_R(k_2)\delta^3(\k_2+\k_3)\delta^3(\k_1+\p_1)\nonumber \\
+& \text{p-cyclic}). 
\end{align}
Putting all the terms together in equation (\ref{effthreepointfun}), we have
\begin{align}\label{finalnongauss}
\langle C|\hat{\R}^{phy}_{\k_1}(\tau)\hat{\R}^{phy}_{\k_2}&(\tau)\hat{\R}^{phy}_{\k_3}(\tau)|C\rangle=(2 \pi)^3 \delta^3(\k_1+\k_2+\k_3)P_R(k_2)P_R(k_1)\nonumber \\
&\times \left[\frac{1}{2}\left(3\epsilon -2\eta+\epsilon\frac{k_1^2+k_2^2}{k_3^2}\right)+\frac{4\epsilon }{ (k_1+k_2+k_3)}\frac{k_1^2k_2^2}{k_3^3}\right]+ \text{cyclic perm}.
\end{align}


\end{document}